\documentclass[1p,sort&compress]{elsarticle}
\usepackage{amsmath,amssymb}

\newcommand{\Tr}{\mathrm{Tr}}%
\newcommand{\hdelta}[1]{{\hat{\delta}}_{#1}}
\def\<#1>{\langle #1 \rangle}%

\usepackage{color}

\begin{document}

\begin{frontmatter}
\title{Derivation of density operators for generalized entropies with quantum analysis}
\journal{}
\author{Masamichi Ishihara\corref{cor}}
\cortext[cor]{Corresponding author. Tel.: +81 24 932 4848; Fax: +81 24 933 6748.}
\ead{m\_isihar@koriyama-kgc.ac.jp}
\address{Department of Human Life Studies, Koriyama Women's University, Koriyama, 963-8503, Japan}
\begin{abstract}
We gave a simple derivation of density operator with the quantum analysis.
We dealt with the functional of a density operator, and applied maximum entropy principle.
We obtained easily the density operators for the Tsallis entropy and R\'enyi entropy with the $q$-expectation value (escort average), 
and also obtained easily the density operators for the Boltzmann-Gibbs entropy and the Burg entropy with the conventional expectation value. 
The quantum analysis works effectively in the calculation of the variation of the functional which includes trace.
\end{abstract}
\end{frontmatter}

\section{Introduction}
Generalized entropies such as Tsallis entropy \cite{Tsallis1998, Book:Tsallis, Bercher2011, Pougaza:AIPConf:2011} 
and R\'enyi entropy \cite{Bercher2011, Pougaza:AIPConf:2011, Abe-PRE94}
are used to study various phenomena in many branches of science. 
It is well-known that the Boltzmann-Gibbs entropy is maximized for the equilibrium state.  
The Burg entropy \cite{Pougaza:AIPConf:2011, Silverstein:Conf:1988, Alexander2003, Mansoury:AMS:2008, Singh:Env.Process:2015}
is also used in researches, such as 
the estimation of autocorrelation\cite{Silverstein:Conf:1988},  streamflow \cite{Singh:Env.Process:2015}, etc. 
It is assumed that the value of an adequate entropy is extremum 
for an achieved state in various systems, 
like the Boltzmann-Gibbs entropy for the thermal equilibrium state. 
The entropy plays an essential role in systems.

The density operator $\rho$ is a basic tool to calculate quantities in quantum statistics. 
The form of the operator is often derived in the maximum entropy principle (MEP) with constraints.
The definition of the entropy and the constraints play essential roles in the MEP. 
The entropy is often defined with trace. 
The functional of the density operator is used to determine the form of the density operator in the MEP.  

The difference between the functional of the density operator with small deviation $\rho+\varepsilon (\delta\rho)$ and 
the functional of $\rho$ is calculated in the MEP.
The density operator, $\rho+\varepsilon (\delta\rho)$, is not diagonal even when $\rho$ is diagonal, 
and $\rho$ and $\delta \rho$ does not commute in general. 
Therefore, the calculation of the difference is not always easy.

The method of the calculation has been developed, and one of them is named quantum analysis 
\cite{Suzuki-commun-math, Suzuki:Book:QuantAnalysis, Suzuki-Review-MathPhys, Suzuki:JMathPhys, Suzuki-progress, Suzuki-IJMPC10}. 
For example, the Taylor expansion has been developed in the analysis, 
and the expansion is similar to the conventional Taylor expansion.
As applications, this analysis was applied to the nonequilibrium response \cite{Suzuki-progress} 
and quantum correlation identites \cite{Suzuki-IJMPB16, Suzuki-IJMPC10}. 
This analysis may simplify the calculation, and the analysis is useful in quantum statistics.

The purpose of this paper is to derive the density operator easily with quantum analysis in the MEP. 
We derive the density operator for the Tsallis entropy and  the R\'enyi entropy 
with the normalized $q$-expectation value (escort average) by applying the quantum analysis.  
For comparison, we derive the density operator for the Boltzmann-Gibbs entropy with the conventional expectation value.
In addition, we derive the density operator for the Burg entropy with the conventional expectation value.
It is understood that the density operator is easily derived with quantum analysis. 

This paper is organized as follows.
In Sec.~\ref{sec:derivation}, 
we give the variation of the functional of a density operator in the quantum analysis, where the functional includes trace.
In Sec.~\ref{sec:density-op:some-entropies}, 
we derive the density operators for the entropies in the MEP.
Sec.~\ref{sec:conclusion} is assigned for conclusion.

\section{Basics of quantum analysis and variations of functionals}
\label{sec:derivation}
We start with the following differential $df(A)$ \cite{Suzuki-commun-math, Suzuki:Book:QuantAnalysis, Suzuki-Review-MathPhys}:
\begin{align}
df(A) = \lim_{\varepsilon \rightarrow 0} \frac{f(A + \varepsilon dA) - f(A)}{\varepsilon} , 
\end{align}
where the operators $A$ and $dA$ do not commute in general.
The above differential is represented symbolically as follows.
\begin{align}
df(A) = \frac{df(A)}{dA} dA .
\label{def:dfdA}
\end{align}
The function $\frac{df(A)}{dA}$ is a hyperoperator, which maps an operator $dA$ to $df(A)$.
The function $\frac{df(A)}{dA}$ is named quantum derivative in the quantum analysis.
The hyperoperator $L_A$ is defined as the left multiplication to the operator: $L_A B :=  AB$.
The inner derivation $\hdelta{A}$ is defined by 
\begin{align}
\hdelta{A} B := [A, B] = AB - BA .
\end{align}
The hyperoperator $L_A$ and the inner derivation $\hdelta{A}$ commute: $L_A \hdelta{A} B  = \hdelta{A} L_A B$.
We do not distinguish between $L_A$ and $A$ throughout this paper as in Ref.~\cite{Suzuki-commun-math} .
With these hyperoperators, the quantum derivative in Eq.~\eqref{def:dfdA} is represented as 
\begin{align}
\frac{df(A)}{dA}  = \int_0^1 dt f^{(1)} (A-t\hdelta{A}) , 
\end{align}
where $f^{(k)}(x)$ is the $k$-th derivative of $f(x)$. 
The variation of the functional $F(A)$ \cite{Suzuki:Book:QuantAnalysis, Suzuki:JMathPhys} is defined by 
\begin{align}
\delta F(A)  = \lim_{\varepsilon \rightarrow 0} \frac{F(A+\varepsilon (\delta A) ) - F(A)} {\varepsilon} . 
\end{align}

We deal with the following cases, where the $f(\rho)$  is a function of an operator $\rho$:
\begin{subequations}
\begin{align}
&F(\rho) = \Tr(f(\rho)) , \\
&G(\rho, A) = \frac{\Tr(g(\rho) A)} {\Tr(g(\rho))} . 
\end{align}
\end{subequations}
The variation of $F(\rho)$ is calculated as follows:
\begin{align}
\delta F(\rho) & 
= \lim_{\varepsilon \rightarrow 0} \frac{1}{\varepsilon} 
\Big\{ \Tr \Big( f(\rho+\varepsilon (\delta \rho) ) \Big) - \Tr(f(\rho)) \Big\}
\nonumber \\ &
= \lim_{\varepsilon \rightarrow 0} \frac{1}{\varepsilon} 
\Big\{ \Tr \Big( f(\rho) + \frac{d f(\rho)}{d\rho} (\varepsilon (\delta \rho) ) + O(\varepsilon^2) \Big)  - \Tr(f(\rho)) \Big\}
\nonumber \\ &
= \lim_{\varepsilon \rightarrow 0} \frac{1}{\varepsilon} 
\Big\{ \Tr \Big( f(\rho) + \int_0^1 dt f^{(1)}(\rho - t \hdelta{\rho}) \varepsilon (\delta \rho)  + O(\varepsilon^2) \Big) - \Tr(f(\rho)) \Big\}
\nonumber \\ &
= \sum_{k=0}   \frac{(-1)^k}{k!} \Bigg( \int_0^1 dt \ t^k \bigg)  \Tr \Bigg( f^{(k+1)}(\rho) (\hdelta{\rho})^k (\delta \rho) \Bigg)
. 
\end{align}
We focus on the term $\Tr( f^{(k+1)}(\rho) (\hdelta{\rho})^k (\delta \rho) )$ which has $2^k$ terms for a positive integer $k$ ($k \ge 1$).  
The term has the  following form: $\Tr (f^{(k+1)}(\rho) \rho^{k-j} (\delta \rho) \rho^j)$ $(0 \le j \le k)$. 
When the trace is invariant under cyclic permutations, 
we have the following result:
\begin{align}
\Tr (f^{(k+1)}(\rho) \rho^{k-j} (\delta \rho) \rho^j) = \Tr (\rho^j f^{(k+1)}(\rho) \rho^{k-j} (\delta \rho) )= \Tr (f^{(k+1)}(\rho) \rho^k (\delta \rho) ) . 
\end{align}
The number of the terms with plus sign is equal to that with minus sign when $k$ is fixed. 
Therefore, the term $\Tr ( f^{(k+1)}(\rho) (\hdelta{\rho})^k (\delta \rho) )$ is equal to zero for $k \ge 1$. 
We obtain
\begin{align}
\delta F(\rho)  =  \Tr \Big( f^{(1)}(\rho) (\delta \rho) \Big). 
\label{eqn:deltaF}
\end{align}
In the same way, the variation of $G(\rho, A)$ under the condition $[\rho, A]=0$ is given by 
\begin{align}
\delta G(\rho, A) = 
\frac{\Tr(g^{(1)}(\rho) (\delta \rho) A)}{\Tr(g(\rho))}
- 
\left( \frac{\Tr(g(\rho) A)}{\Tr(g(\rho))} \right)
\left( \frac{\Tr(g^{(1)}(\rho) (\delta \rho) )}{\Tr(g(\rho))} \right)
.
\label{eqn:deltaG}
\end{align}

\section{Derivation of density operators for some entropies in the maximum entropy principle}
\label{sec:density-op:some-entropies}
In the following calculations, we attempt to obtain the density operator with the quantum analysis in the MEP. 
The density operators for the Tsallis entropy and the R\'enyi entropy with the normalized $q$-expectation value are derived. 
For comparison, the density operator for the Boltzmann-Gibbs entropy with the conventional expectation value is derived. 
The density operator for the Burg entropy with the conventional expectation value is also derived.

\subsection{Tsallis entropy with the normalized $q$-expectation value}
The entropy and the normalized $q$-expectation value  in the Tsallis nonextensive statistics are defined by 
\begin{subequations}
\begin{align}
& S_{\mathrm{T},q}(\rho) = \frac{1-\Tr(\rho^q)}{q-1}, \\
& \<A>_q = \frac{\Tr(\rho^q A)}{\Tr (\rho^q)} ,
\end{align}
\end{subequations}
where $\rho$ is the density operator. 
The functional $I_{\mathrm{T}}(\rho)$ is defined by 
\begin{align}
I_{\mathrm{T}}(\rho) := S_{\mathrm{T},q}(\rho) -\alpha (\Tr(\rho) -1) - \beta (\<H>_q - E) , 
\end{align}
where $\alpha$ and $\beta$ are Lagrange multipliers, and $H$ is the Hamiltonian. 
We attempt to obtain $\rho$ as a function of $H$ by imposing the condition $\delta I_T(\rho)=0$.
The commutation relation $[\rho, H]=0$ is satisfied, 
and we easily obtain $\delta I_T(\rho)$ using Eqs.~\eqref{eqn:deltaF} and \eqref{eqn:deltaG} with $f(\rho)=g(\rho) = \rho^q$:
\begin{align}
\delta I_{\mathrm{T}}(\rho) 
= \Tr \Bigg\{  \Bigg[ \Bigg( \frac{q}{(1-q)} \Bigg) \rho^{q-1} - \alpha - q \Bigg( \frac{\beta}{\Tr (\rho^q)} \Bigg) (H-\<H>_q) \rho^{q-1} \Bigg] (\delta \rho) \Bigg\}
.
\end{align}

The requirement $\delta I_{\mathrm{T}}(\rho) = 0$ gives 
\begin{align}
\Bigg( \frac{q}{(1-q)} \Bigg) \rho^{q-1} - \alpha - q \Bigg( \frac{\beta}{\Tr (\rho^q)} \Bigg) (H-\<H>_q) \rho^{q-1} = 0 .
\end{align}
We finally obtain the density operator with the condition $\Tr \rho =1$:
\begin{subequations}
\begin{align}
&\rho = \frac{1}{Z_{\mathrm{T}}}\Bigg[ 1 - (1-q) \Bigg(\frac{\beta}{\Tr (\rho^q)} \Bigg) (H-\<H>_q) \Bigg]^{\frac{1}{1-q}}  ,\\ 
& Z_{\mathrm{T}} =  \Tr \Bigg\{ \Bigg[ 1 - (1-q) \Bigg(\frac{\beta}{\Tr (\rho^q)} \Bigg) (H-\<H>_q) \Bigg]^{\frac{1}{1-q}} \Bigg\} . \nonumber 
\end{align}
\end{subequations}
This density operator for the Tsallis entropy is well-known \cite{Book:Tsallis, Aragao-PhysicaA2003},  
and has been applied to various phenomena.

\subsection{R\'enyi entropy with the normalized $q$-expectation value}
The R\'enyi entropy is defined by  
\begin{align}
S_{\mathrm{R},q}(\rho)  = \frac{\ln \Tr (\rho^q)}{1-q} . 
\end{align}
The variation of $S_{\mathrm{R},q}$ is calculated as follows:
\begin{align}
S_{\mathrm{R},q}(\rho+\varepsilon (\delta \rho)) 
&= \Bigg( \frac{1}{1-q} \Bigg) \ln \Tr \Bigg\{ \rho^q + \int_0^1 \ dt\ q (\rho-t \hdelta{\rho})^{q-1} \left( \varepsilon (\delta \rho) \right) + O(\varepsilon^2) \Bigg\}
\nonumber \\
&= \Bigg(\frac{1}{1-q} \Bigg) \ln \Tr \Bigg\{ \rho^q + q \int_0^1 \ dt \rho^{q-1} \left( \varepsilon (\delta \rho) \right) + O(\varepsilon^2) \Bigg\}
\nonumber \\
&= \Bigg(\frac{1}{1-q} \Bigg)\ln \Bigg\{ 
(\Tr \rho^q ) \Bigg( 1 +  \varepsilon q \frac{\Tr \Big( \rho^{q-1} (\delta \rho) \Big) }{\Tr \rho^q}  + O(\varepsilon^2) \Bigg)
\Bigg\}
\nonumber \\
&= \Bigg(\frac{1}{1-q} \Bigg)
\Bigg\{ \ln (\Tr \rho^q ) + \varepsilon q \frac{\Tr \Big( \rho^{q-1} (\delta \rho) \Big) }{\Tr \rho^q}  + O(\varepsilon^2) \Bigg\} 
. 
\end{align}
Therefore, we have
\begin{align}
\delta S_{\mathrm{R},q}(\rho)  = \Bigg( \frac{q}{1-q} \Bigg)  \frac{\Tr \Big( \rho^{q-1} (\delta \rho) \Big) }{\Tr \rho^q} .
\end{align}

The functional $I_{\mathrm{R}}(\rho)$ with the constraints given by the normalized $q$-expectation value is 
\begin{align}
& I_{\mathrm{R}}(\rho)  = S_{\mathrm{R},q}(\rho) - \alpha(\Tr \rho -1) - \beta ( \<H>_q  -  E ) .
\end{align}
The requirement $\delta  I_{\mathrm{R}}(\rho) = 0$ using Eq.~\eqref{eqn:deltaG} with $g(\rho) = \rho^q$ gives 
\begin{align}
\Tr \Bigg\{
\Bigg[ \Bigg( \frac{q}{1-q} \Bigg)  \frac{\rho^{q-1}}{\Tr \rho^q} - \alpha - q \Bigg( \frac{\beta}{\Tr (\rho^q)} \Bigg) (H-\<H>_q) \rho^{q-1} \Bigg] (\delta \rho) 
\Bigg\} = 0 .
\end{align}
We finally obtain the density operator for R\'enyi entropy with the normalized $q$-expectation value:
\begin{align}
&\rho = \frac{1}{Z_{\mathrm{R}}} \Big[ 1 - (1-q) \beta (H-\<H>_q) \Big]^{\frac{1}{(1-q)}} , \\
&Z_{\mathrm{R}} = \Tr \Big[ 1 - (1-q) \beta (H-\<H>_q) \Big]^{\frac{1}{(1-q)}} . \nonumber 
\end{align}
The density operator for the R\'enyi entropy resembles the density operator  for the Tsallis entropy.  
The form of the above density operator agrees with the form of the probability density obtained in Ref.~\citep{Bercher2011}

\subsection{Boltzmann-Gibbs entropy with the conventional expectation value}
We also attempt to derive the density operator in the Boltzmann-Gibbs statistics for comparison.
The functional $I_{\mathrm{BG}} (\rho)$ is 
\begin{subequations}
\begin{align}
& I_{\mathrm{BG}}(\rho)  = S_{\mathrm{BG}}(\rho)  - \alpha(\Tr \rho -1) - \beta ( \<H> -  E ) , \\
& S_{\mathrm{BG}}(\rho) = - \Tr (\rho \ln \rho) , 
\end{align}
\end{subequations}
where $\<H> = \Tr (\rho H)/\Tr \rho$. 
The requirement $\delta I_{\mathrm{BG}}(\rho)=0$ 
using Eq.~\eqref{eqn:deltaF} with  $f(\rho)=-\rho \ln \rho$ and Eq.~\eqref{eqn:deltaG} with $g(\rho) = \rho$ gives 
\begin{align}
\delta I_{\mathrm{BG}} = \Tr \Bigg\{ \Bigg[ -(\ln \rho + 1) -\alpha - \frac{\beta}{\Tr \rho} (H-\<H>) \Bigg] (\delta \rho)\Bigg\} = 0, 
\end{align}
We obtain 
\begin{align}
\rho = \exp(-\alpha-1) \exp\Big(-\frac{\beta}{\Tr \rho} (H-\<H>)\Big) .
\end{align}
The requirement $\Tr \rho = 1$ leads to 
\begin{subequations}
\begin{align}
& \rho = \frac{1}{Z_{\mathrm{BG}}} \exp\Big(-\beta(H-\<H>)\Big), \\ 
& Z_{\mathrm{BG}} = \Tr \Bigg[ \exp\Big(-\beta(H-\<H>)\Big) \Bigg] . 
\end{align}
\end{subequations}
The above density operator is well-known in the Boltzmann-Gibbs statistics, and is equivalent to $\exp (-\beta H)/\Tr(\exp(-\beta H))$. 
The expectation value $\<H>$ appears in the above density operator, as in the density operator for the Tsallis entropy. 
This implies that the appearance of $\<H>$ in the density operator is natural.

\subsection{Burg entropy with the conventional expectation value}
We attempt to find the density operator for the entropy analogous to Burg entropy.
The Burg entropy is defined with the probability $p_i$ of a state $i$: 
\begin{align}
S_{\mathrm{Bu}}^{(\mathrm{cl})} = \sum_i \ln  p_i. 
\label{S:burg}
\end{align}
To use the Burg entropy in quantum systems, we use the following form of the entropy with the density operator:
\begin{align}
S_{\mathrm{Bu}}(\rho) = \Tr \ln \rho . 
\end{align}

The variation of $S_{\mathrm{Bu}}(\rho)$ is given using Eq.~\eqref{eqn:deltaF} with  $f(\rho)= \ln \rho$:
\begin{align}
\delta S_{\mathrm{Bu}}(\rho) = \Tr \bigg[ \bigg( \frac{1}{\rho} \bigg)  (\delta \rho) \bigg]  . 
\end{align}

The functional $I_{\mathrm{Bu}}(\rho)$ is given by 
\begin{align}
I_{\mathrm{Bu}}(\rho)  = S_{\mathrm{Bu}}(\rho) - \alpha(\Tr \rho -1) - \beta ( \<H> -  E ) . 
\end{align}
Therefore, the variation of the functional  $\delta  I_{\mathrm{Bu}}(\rho)$ using Eq.~\eqref{eqn:deltaG} with $g(\rho) = \rho$ is 
\begin{align}
\delta I_{\mathrm{Bu}}(\rho) = \Tr \Bigg\{ \Bigg[ \frac{1}{\rho} -\alpha - \frac{\beta}{\Tr \rho} (H-\<H>) \Bigg] (\delta \rho)\Bigg\} .
\end{align}
The requirement $\delta  I_{\mathrm{Bu}}(\rho) =0$ gives
\begin{align}
\frac{1}{\rho} = \alpha + \frac{\beta}{\Tr \rho} (H-\<H>) = \alpha \bigg( 1 + \frac{\tilde{\beta}}{\Tr \rho} (H-\<H>) \bigg)
, \qquad \tilde{\beta} = \beta/\alpha, 
\end{align}
The constraint $\Tr \rho = 1$ gives 
\begin{subequations}
\begin{align}
& \rho = \frac{1}{Z_{\mathrm{Bu}}} \bigg( \frac{1}{1+ \tilde{\beta} (H-\<H>)} \bigg) , 
\label{density-op:burg:qm}\\
& Z_{\mathrm{Bu}} = \Tr \bigg( \frac{1}{1+ \tilde{\beta} (H-\<H>)} \bigg) .
\end{align}
\end{subequations}
The above density operator agrees with the probability obtained previously \cite{Pougaza:AIPConf:2011, Mansoury:AMS:2008} for the Burg entropy.
The density operator, Eq.~\eqref{density-op:burg:qm}, should be carefully treated, 
as the Burg entropy is not defined when the probability $p_i$ is zero for a certain state $i$.

\section{Conclusion} 
\label{sec:conclusion}
We attempted to find density operators for some entropies with the quantum analysis. 
We derived the density operators for the Tsallis entropy and the R\'enyi entropy with the normalized $q$-expectation value (escort average), 
and derived the density operator for the Boltzmann-Gibbs entropy with the conventional expectation value, for comparison.   
We also derived the density operator for the Burg entropy.
These density operators were derived easily with the quantum analysis. 

The quantum analysis works well to derive the density operators for various entropies with constraints in the maximum entropy principle.
The quantum analysis is useful in quantum statistics,  
because trace often appears and the inner derivation $\hdelta{A}$ which acts as $\hdelta{A} B \equiv AB-BA$ works effectively. 
It is possible to extend the current method to the grand canonical ensemble.

The quantum analysis gives the method of the calculation in quantum systems.  
Especially, the analysis will work in calculations of trace terms,
because the quantum derivative is represented with $\hdelta{A}$. 

The author hopes that the present study is a cue to calculate quantities with the quantum analysis in quantum statistics.


\end{document}